\documentclass[aps,twocolumn,nofootinbib]{revtex4}

\usepackage{graphicx}
\usepackage{dcolumn}
\usepackage{bm}

\usepackage{amssymb,amsfonts}
\usepackage{epsfig}
\usepackage{epstopdf}
\usepackage[all]{xy}
\usepackage{amsthm}
\usepackage{dcolumn}
\usepackage{hyperref}
\usepackage{url}
\usepackage{dsfont}

\newcommand{\be}{\begin{equation}}
\newcommand{\ee}{\end{equation}}
\newcommand{\bea}{\begin{eqnarray}}

\newcommand{\eea}{\end{eqnarray}}

\def\inbar{\,\vrule height1.5ex width.4pt depth0pt}
\def\IR{\relax{\rm I\kern-.18em R}}
\def\IC{\relax\hbox{$\inbar\kern-.3em{\rm C}$}}

\begin{document}

\title{`Hidden' symmetry of linearized gravity in de Sitter space}

\author{Hamed Pejhan$^{1}$\footnote{pejhan@zjut.edu.cn}}

\author{Surena Rahbardehghan$^2$\footnote{sur.rahbardehghan.yrec@iauctb.ac.ir}}

\author{Mohammad Enayati$^1$}

\author{Kazuharu Bamba$^{3}$\footnote{bamba@sss.fukushima-u.ac.jp}}

\author{Anzhong Wang$^{1,4}$\footnote{Anzhong-Wang@baylor.edu}}

\affiliation{$^1$Institute for Theoretical Physics and Cosmology, Zhejiang University of Technology, Hangzhou 310032, China}

\affiliation{$^2$Department of Physics, Science and Research Branch, Azad University, Tehran 1477893855, Iran}

\affiliation{$^3$Division of Human Support System, Faculty of Symbiotic Systems Science, Fukushima University, Fukushima 960-1296, Japan}

\affiliation{$^4$GCAP-CASPER, Physics Department, Baylor University, Waco, TX 76798-7316, USA}

\begin{abstract}
We demonstrate that the linearized Einstein gravity in de Sitter (dS) spacetime besides the evident symmetries also possesses the additional (local) symmetry $h_{\mu\nu}\rightarrow h_{\mu\nu} + {\cal E}_{\mu\nu}\chi$, where ${\cal E}_{\mu\nu}$ is a spin-two projector tensor and $\chi$ is an arbitrary constant function. We argue that an anomalous symmetry associated with this hitherto `hidden' property of the existing physics is indeed at the origin of `dS breaking' in linearized quantum gravity.
\end{abstract}
\maketitle

\section{Introduction}
Recent three decades have witnessed a proliferation of conflicting approaches to dS quantum gravity, because of the long-standing puzzle of dS symmetry breaking for quantum field theory of gravity in dS spacetime (see \cite{Antoniadis2007} and references therein). Together with the quantization of the dS minimally coupled scalar field, they set up a doublet of cases where dS quantum field theory would lead to such difficulties. The case of the minimally coupled scalar field was analyzed by Allen and Folacci \cite{AF} and their results seemed definitive: dS invariance was broken and infrared divergences were present. The case of the graviton field, however, is more complicated among other things due to the presence of the local invariance, contrary to the scalar case, and it is still a source of contention in the literature: while the mathematical physics community maintain that there is no physical breaking of dS invariance, the particle physics community argue that gravitons inherit the dS breaking long recognized for the minimally coupled scalar field (see \cite{Antoniadis2007,MorrisonarXiv} and references therein). It is therefore pertinent to develop theoretical benchmarks that will allow us to select the physically relevant approaches, and to subsequently use observational constraints in order to single out the few candidates that are actually the viable ones.

From the very beginning of this scientific dispute, a firm reasoning in favor of covariant quantization of the graviton field in the natural dS vacuum state (the Bunch-Davies vacuum) was put forward in Ref. \cite{higuchi}, and during recent three decades, it has been dynamically subject to scrutiny in a number of works (see, for instance, \cite{Bernar,HH Higuchi'',HH Frob,HH Bernar,Fewster,Faizal&Higuchi,HH Higuchi',Marolf&Morrison,HH Higuchi,higuchi1,HH Hartle,HH Allen,HH Marolf,FrobJCAP}). Let us make the idea lying behind of this reasoning explicit, using the so-called conformal (global) coordinates,
\begin{eqnarray}
&x=(x^0=H^{-1}\tan\rho, (H\cos\rho)^{-1} u), \rho\in]\frac{-\pi}{2},\frac{\pi}{2}[, u\in S^3.& \nonumber
\end{eqnarray}
Basically, in Refs. \cite{higuchi,Bernar,HH Higuchi'',HH Frob,HH Bernar,Fewster,Faizal&Higuchi,HH Higuchi',Marolf&Morrison,HH Higuchi,higuchi1,HH Hartle,HH Allen,HH Marolf,FrobJCAP}, the physical graviton modes are respected as the transverse-traceless second-rank symmetric tensor spherical harmonics on the three-spheres, symbolized here by $h_{\mu\nu}^{(Llm)}(\rho,u)$, with $h_{0\mu}=0$ (the synchronous condition), $L=2,3,...$, $0\leq l\leq L$ and $-l\leq m\leq l$. It is indeed argued that, by imposing the transverse-traceless-synchronous (TTS) gauge conditions, each irreducible component of the representation of the dS group $SO_0(1,4)$ given by linearized gravity (i.e., $\Pi_{2,2}^{+}\oplus \Pi_{2,2}^{-}$ in the Dixmier's notation\cite{Dixmier}) is formed by this set of solutions \cite{higuchi}. This argument seems to entail making sense of the Bunch-Davies vacuum as the unique possibility for a dS-invariant dynamical gravitons state, and that implies somehow avoiding the appearance of the infrared divergences and the associated symmetry breaking in the theory.

The critical point that should be noted here is that ``\emph{It is not possible to find $h_{\mu\nu}$ satisfying the TTS gauge conditions in this form (i.e. the second-rank symmetric tensor spherical harmonics on the three-spheres) if $L=0$ or $1$}" \cite{Bernar}. Indeed, in this approach to dS quantum gravity, because the normalization factor breaks down at $L=0$ corresponding to the graviton zero-frequency mode, this mode is not considered in the set of solutions based upon which the Fock space and the vacuum state are constructed. It is shown that the value $L=1$ is not allowed either. See more mathematical details in Refs. \cite{higuchi,Bernar,HH Higuchi'',HH Frob,HH Bernar,Fewster,Faizal&Higuchi,HH Higuchi',Marolf&Morrison,HH Higuchi,higuchi1,HH Hartle,HH Allen,HH Marolf,FrobJCAP}.

In this paper, we sketch the parallel steps in a coordinate-independent approach based on ambient space notations. We show that the linearized Einstein gravity in dS spacetime besides the evident symmetries also possesses a hitherto `hidden' local symmetry. This local symmetry of the classical theory, which is reflected by the subspace generated by $L=0$ mode in the set of solutions, reveals that the above appealing picture of dS quantum gravity in the TTS gauge ($L\geq2$) does not transform correctly under the whole symmetries of the theory, and therefore, does not lead to covariant results. We also prove that this local symmetry becomes anomalous in the quantized theory. This anomaly makes the theory inconsistent and must be canceled at all costs. The cancelation imposes strong restrictions on the theory which explicitly violates the covariance condition of the usual canonical quantization of the graviton field in dS spacetime.

\section{Covariant formulation}
We start our discussion by reminding the standard procedure to study linear perturbations of Einstein gravity around the dS metric. We denote the metric perturbation by a symmetric tensor $h_{\mu\nu}$ propagating on a dS fixed background $\hat{g}_{\mu\nu}$, $g_{\mu\nu}=\hat{g}_{\mu\nu}+h_{\mu\nu}$. This reparametrization invariance is translated at the linear level as the following gauge invariance
\begin{eqnarray}\label{GT}
h_{\mu\nu} \rightarrow h_{\mu\nu} + 2\nabla_{(\mu}\zeta_{\nu)},
\end{eqnarray}
where $\zeta_{\mu}$ is an arbitrary vector field and $\nabla_\mu$ is the dS covariant derivative. In this framework, the linearized Einstein equation of motion in dS spacetime takes the form
\begin{eqnarray} \label{2}
&(\Box_H + 2H^2)h_{\mu\nu} - (\Box_H - H^2)\hat{g}_{\mu\nu}h' - 2\nabla_{(\mu}\nabla^\rho h_{\nu)\rho} &\nonumber \\
&+ \hat{g}_{\mu\nu}\nabla^\lambda\nabla^\rho h_{\lambda\rho} + \nabla_\mu\nabla_\nu h'=0,&
\end{eqnarray}
where $H$ is the Hubble constant, $\Box_H= \hat{g}_{\mu\nu}\nabla^\mu\nabla^\nu$ is the Laplace-Beltrami operator, and $h'=\hat{g}^{\mu\nu}h_{\mu\nu}$ is the trace of $h_{\mu\nu}$ (all tensor indices are raised and lowered by the background metric). We will partially fix the gauge by using the Lorenz gauge condition, defined by $\nabla^\mu h_{\mu\nu} = {e}\nabla_\nu h'$, with $e=1/2$.

\subsection{Isometric embedding}
For the sake of argument, it is useful to consider a more convenient set of coordinates, defined by the isometric embedding of the $3+1$-dimensional dS space in the ${4+1}$-dimensional Minkowski space (${{\mathbb{M}}}_5$) as the ambient space,
\begin{eqnarray}
&{M_{H}}  = \{ x \in \mathbb{M}_5 ;\; x^2 ={\eta}_{\alpha\beta} {x^{\alpha}} {x^\beta} =  -H^{-2}\},&\nonumber\\
&\alpha,\beta=0,1,2,3,4, \;\; \eta_{\alpha\beta}= \mbox{diag}(1,-1,-1,-1,-1),&\nonumber
\end{eqnarray}
with the ambient coordinates notations $x\equiv (x^0, \overrightarrow{x}, x^4)$. The dS metric would be the induced metric on the dS hyperboloid
\begin{eqnarray}
ds^2= \eta_{\alpha\beta}dx^{\alpha}dx^{\beta}|_{x^2=-H^{-2}} = {\hat{g}}_{\mu\nu} dX^{\mu}dX^{\nu}, \nonumber
\end{eqnarray}
where $\mu,\nu=0,1,2,3$ and $X^{\mu}$'s are the four local spacetime coordinates for the dS hyperboloid.

Utilizing the ambient space formalism is justified for two reasons. First, this embedding is a purely geometrical construction and hence has no effect on the dynamics, which always takes place in $3+1$-dimensions. As a matter of fact, in the ambient space notations, a tensor field ${\cal K}$ is a homogeneous function of the ${{\mathbb{M}}}_5$-variables $x^\alpha$, so that its degree of homogeneity is fixed $x_{\alpha}(\partial/\partial x_{\alpha}){\cal K}\equiv x\cdot\partial {\cal K}=\varrho {\cal K}$, with $\varrho\in {\mathbb{R}}$, and we set $\varrho = 0$ to have correspondence between $\Box_H\equiv\nabla_\mu\nabla^\mu$ on dS space and $\Box_5\equiv \partial^2$ on ${{\mathbb{M}}}_5$. Moreover, to ensure that ${\cal K}$ lies in the dS tangent space, it is constrained to obey the transversality condition $x\cdot {\cal K}=0$. Respecting the significance of the transversality property for dS fields, the transverse projection is defined by $(\mbox{T}{\cal K})_{\alpha_1...\alpha_s}=\theta_{\alpha_1}^{\beta_1}...\theta_{\alpha_s}^{\beta_s}{\cal K}_{\beta_1...\beta_s}$, with $\theta_{\alpha\beta}=\eta_{\alpha\beta}+H^2x_\alpha x_\beta$, which guarantees the transversality in each index. $\theta_{\alpha\beta}$ is indeed the only tensor which corresponds to the metric ${\hat{g}}_{\mu\nu} = x_{\mu}^{\alpha}x_{\nu}^{\beta}\theta_{\alpha\beta}$, with $x_{\mu}^{\alpha}=\partial x^{\alpha}/\partial X^{\mu}$. [Pursuing the same path, any intrinsic tensor field like the graviton field can be locally specified by its transverse counterpart in the ambient space notations, $h_{\mu\nu}(X)= x_{\mu}^{\alpha}x_{\nu}^{\beta}{\cal{K}}_{\alpha\beta}(x(X))$. The covariant derivatives acting on a symmetric, second-rank tensor are also transformed according to $\nabla_\rho\nabla_\lambda h_{\mu\nu} = x_{\rho}^{\alpha}x_{\lambda}^{\beta}x_{\mu}^{\gamma}x_{\nu}^{\sigma}\mbox{T}\bar\partial_\alpha \mbox{T}\bar\partial_\beta {\cal{K}}_{\gamma\sigma}$, in which $\bar\partial \equiv \mbox{T}\partial$ stands for transverse derivative in dS space.]

Second, this way of describing the dS space constitutes the coordinate-independent approach, such that there is a close resemblance with the corresponding description on the Minkowski space and the link with group theory is easily readable. More technically, in the ambient notations, the ten infinitesimal generators $L_{\alpha\beta}^{(r)}$ of the dS group are simply defined by $L_{\alpha\beta}^{(r)} = M_{\alpha\beta} + S_{\alpha\beta}^{(r)}$ \cite{Gazeau2533}, where the orbital part is $M_{\alpha\beta}=-i(x_{\alpha}\partial_{\beta}-x_{\beta}\partial_{\alpha})$ and the spinorial part $S_{\alpha\beta}^{(r)}$ acts on the indices of rank-$r$ tensors as $S_{\alpha\beta}^{(r)}{\cal K}_{\alpha _1 ... \alpha _r} = -i\displaystyle\sum_{i=1}^{r}\Big(\eta_{\alpha\alpha _i}{\cal K}_{\alpha _1...(\alpha _i\rightarrow \beta)...\alpha_r} - (\alpha\leftrightharpoons\beta)\Big)$. The second-order Casimir operator of the dS group $Q_r\equiv -\frac{1}{2}L_{\alpha\beta}^{(r)}L^{{(r)}{\alpha\beta}}$ is fixed on the carrier space of each dS unitary irreducible representation (UIR), and therefore, it can be used to classify the dS UIR's,
\begin{eqnarray}\label{qr}
Q_r=\langle Q_r\rangle \mathbb{I}=[-p(p+1)-(q+1)(q-2)]\mathbb{I},
\end{eqnarray}
with $2p \in\mathbb N$ and $q\in\mathbb C$ (in the Dixmier's notation \cite{Dixmier}). The irreducible representations associated with our study are characterized by the dS discrete series representations $\Pi_{p,q}^{\pm}$, in which label `$q$' has a spin meaning and `$\pm$' stands for the two types of helicity. For $p=q$, these representations correspond to the conformal massless cases. More precisely, the term `\emph{massless}' is used here in reference to the conformal invariance (propagating on the dS light cone). See a more detailed discussion in \cite{levy,barut,bacry}. In this sense, the dS massless spin-2 field is described by the UIR's $\Pi^\pm_{2,2}$, with $Q_{r=2}\equiv Q_2=\langle Q_2\rangle \mathbb{I}=-6\mathbb{I}$, and lies among the solutions of the dS-invariant equation $(Q_2+6){\cal K}=0$, supplemented with the divergencelessness condition $\partial\cdot {\cal K}=0$.

Taking all of the above into consideration, the ambient space counterpart of the linearized Einstein equations of motion (\ref{2}) would be \cite{Fronsdal1979,PejhanII}
\begin{eqnarray}\label{14}
(Q_2+6){\cal K}+D_2\partial_2\cdot {\cal K}=0,
\end{eqnarray}
which can be derived from the following action
\begin{eqnarray}\label{21}
S=\int d\sigma {\cal L}_{i}, \; {\cal L}_{i}=-\frac{1}{2x^2}{\cal K}\cdot\cdot(Q_2+6){\cal K}+\frac{1}{2}(\partial_2\cdot {\cal K})^2,
\end{eqnarray}
where $d\sigma$ is the volume element in dS space, $D_2\equiv H^{-2}{\cal S}(\bar{\partial}-H^2x)$ in which ${\cal S}$ is the symetrizer operator ${\cal S}\vartheta _\alpha \omega _\beta =\vartheta _\alpha \omega _\beta +\vartheta_\beta \omega _\alpha $, $\partial_2\cdot {\cal K}=\partial\cdot {\cal K}-H^2x{\cal K}^\prime -\textstyle\frac{1}{2}\bar{\partial}{\cal K}^\prime$ in which ${\cal K}^\prime$ is the trace of ${\cal K}_{\alpha\beta}$, and `$\cdot\cdot$' is a shortened notation for total contraction. The Lagrangian density (\ref{21}) is invariant under the gauge transformation ${\cal K}\rightarrow {\cal K} + D_2\lambda$, where $\lambda$ is a vector field. This gauge transformation is exactly the ambient space counterpart of (\ref{GT}). The Lorenz gauge condition in ambient space notations takes the form $\partial_2\cdot {\cal K} = (b-1/2)\bar\partial{\cal K}^\prime$, with $b=1/2$. Hence, the gauge fixing can be accomplished by adding the gauge-fixing term ${\cal L}_{gf}=(1/2a)(\partial_2\cdot {\cal K})^2$ to (\ref{21}), where `$a$' is an arbitrary constant. Finally, the field equation derived from ${\cal L}={\cal L}_{i} + {\cal L}_{gf}$ is
\begin{eqnarray}\label{24}
(Q_2+6){\cal K}+cD_2\partial_2\cdot {\cal K}=0,
\end{eqnarray}
where $c=(1+a)/a$ is the gauge-fixing parameter ($c\neq 1$). Obviously, the UIR's $\Pi^\pm_{2,2}$, and hence the corresponding fields, are part of an indecomposable structure issued from the existence of gauge solutions. See a detailed discussion in Ref. \cite{Bamba2}.

\section{The `zero-mode' problem}
The general solution ${\cal K}_{\alpha\beta}$ to the field equation (\ref{24}) can be written as the resulting action of a second-order differential operator (the spin-two projector) on a minimally coupled scalar field (the structure function), ${\cal K}_{\alpha\beta}\equiv {\cal E}_{\alpha\beta}\phi$, as follows \cite{Bamba2}
\begin{eqnarray}\label{=/2/5}
{\cal K}= {\cal K}^{c=\frac{2}{5}} + \frac{c-\frac{2}{5}}{c-1} D_2 (Q_1+6)^{-1} \partial_2\cdot {\cal K}^{c=\frac{2}{5}}, \; c\neq1 \;\;
\end{eqnarray}
with
\begin{eqnarray} \label{=2/5}
&{\cal K}^{c=\frac{2}{5}}= \Big(-\frac{2}{3}\theta Z\cdot K +{\cal S}\bar{Z}K + \frac{1}{27} H^2 D_2D_1 Z\cdot K& \nonumber\\
& + \frac{1}{3 }H^2 D_2 x\cdot Z K \Big) - \frac{1}{15} D_2\partial_2\cdot {\cal K}^{c=\frac{2}{5}},&
\end{eqnarray}
and
\begin{eqnarray}\label{K}
K = \Big(\bar{Z^\prime}-\frac{1}{2}D_1(Z^\prime\cdot\bar{\partial}+2H^2x\cdot Z^\prime)\Big)\phi,
\end{eqnarray}
\begin{eqnarray}\label{mmcsf}
Q_0\phi = -H^{-2}\Box_H \phi =0,
\end{eqnarray}
in which $Z$ and $Z^\prime$ are two constant five-vectors ($\bar{Z}\equiv \mbox{T}Z$). Note that, the second term in (\ref{=/2/5}) is responsible for the appearance of logarithmic divergences in the theory which can be eliminated by setting $c=2/5$. This specific value of $c$ leads to the simplest structure for the involved indecomposable representation of the dS group (see \cite{Gazeau1984, Gazeau1985g} and references cited therein). In what follows, however, we will work with an arbitrary value for $c$ to retain the generality of our discussion.

The minimally coupled scalar field, corresponding to the lowest case in the scalar discrete series representation $\Pi_{p=1,0}$,\footnote{The scalar representations $\Pi_{p,0}$ in the discrete series are characterized by $Q_0=-(p-1)(p+2)\mathbb{I}$, with $p=1,2,...\;$ \cite{Dixmier}.} can be identified by the so-called coordinate-independent dS plane waves as follows
\begin{eqnarray}\label{dSpw}
\phi(x)=(Hx\cdot\xi)^{\kappa},\;\;\;\;\;    \kappa = - p - 2 = -3,
\end{eqnarray}
where this 5-vector $\xi$ lies on the null cone in ${{\mathbb{M}}}_5$, $\xi = (\xi^0,\boldsymbol\xi) \in {\cal C} = \{ \xi \in {{\mathbb{M}}}_5 ; \; \xi^2 = 0 \}$. [Note that, the case $\kappa = p - 1 = 0$ is referred to as the trivial solution for the minimally coupled scalar field.] The solution (\ref{dSpw}) is defined on connected open subsets of $M_H$ such that $x\cdot\xi\neq 0$ (see \cite{BGM,BM} for details).

Substituting $\boldsymbol\xi = \|\boldsymbol\xi\|v\in{\mathbb{M}}_4$, $v\in S^3$, and $|\xi^0| = \|\boldsymbol\xi\|$, we obtain
\begin{eqnarray}
Hx\cdot \xi = (\tan \rho)\xi^0 - \frac{1}{\cos \rho}u\cdot \boldsymbol\xi = \frac{\xi^0e^{i\rho}}{2i\cos \rho} (1 + z^2 - 2zt),\nonumber
\end{eqnarray}
with $z =  ie^{-i\rho}\mbox{sgn}\;\xi^0$ and $t = u\cdot v \equiv \cos \varpi$. Then, considering the generating function for Gegenbauer polynomials, $(1 + z^2 - 2zt)^{-\lambda} = \sum_{n=0}^\infty z^n C_n^\lambda (t)$, with $|z|<1$, we have the following expansion ($\lambda=-\kappa $)
\begin{eqnarray}\label{sol2}
(Hx\cdot \xi)^{\kappa}  =  \Big[ \Big( \frac{\xi^0e^{i\rho}}{2i\cos \rho} \Big)^{\kappa} \displaystyle\sum\limits_{n=0}^\infty z^nC_n^{-\kappa}(t) \Big], \;\; \Re\kappa <\frac{1}{2}.
\end{eqnarray}
This formula is not valid in terms of functions since $|z| = 1$. Nonetheless, the convergence is ensured if we give a negative imaginary part to the angle $\rho$. Consequently, we extend ambient coordinates to the forward tube \cite{BM},
$$ {\cal{T}}^+ = \{ \mathbb{M}_5 - i\bar{V}_5^+ \cap M_H^{\mathbb{C}} \},\; \bar{V}_5^+ = \{ x\in \mathbb{M}_5: x^2\geq0,x^0>0 \}.$$

Now, the dS plane waves (\ref{sol2}), using two expansion formulas involving Gegenbauer polynomials and normalized hyperspherical harmonics on $S^3$ and after putting $\kappa =-3$, can be written as \cite{MVFgazeau}
\begin{eqnarray}\label{expan}
(Hx\cdot\xi)^{-3}=2\pi^2\sum_{Llm}\Phi^{-3}_{Llm}(\xi^0)^{-3}(\mbox{sgn}\;\xi^0)^LY_{Llm}^\ast(v),
\end{eqnarray}
with
\begin{eqnarray}\label{SOF}
\Phi_{Llm}^{-3}(x) =\frac{i^{L+3}e^{-i(L+3)\rho}}{(2\cos \rho)^{-3}}p_L^{3}(-e^{-2i\rho})Y_{Llm}(u),
\end{eqnarray}
in which $Y_{Llm}$ stands for the $S^3$ hyperspherical harmonics, with $(L,l,m)\in \mathbb{N}\times\mathbb{N}\times\mathbb{Z}$, $0\leq l\leq L$ and $-l\leq m \leq l$, and
\begin{eqnarray}
p_L^3(z^2) = \frac{\Gamma(L+3)}{(L+1)!\Gamma(3)} \,_2F_1(L+3,2;L+2;z^2).
\end{eqnarray}
The hyperspherical harmonics are linearly independent, therefore the functions $\Phi_{Llm}^{-3}$ would be solutions to (\ref{mmcsf}) when adopting the appropriate separation of variables. With respect to the orthonormality of $Y_{Llm}$'s, then we have
\begin{eqnarray}\label{Fourier}
\Phi^{-3}_{Llm}(x) = \frac{(\mbox{sgn}\xi^0)^L}{2\pi^2 (\xi^0)^{-3}} \int_{S^3} d\sigma(v) (Hx\cdot\xi)^{-3}Y_{Llm}(v).\;
\end{eqnarray}
Note that, $\Phi^{-3}_{Llm}(x)$ are well defined, and are infinitely differentiable in the conformal coordinates $(\rho,u)$ \cite{BM}.

The above formulas present the `spherical' modes in dS spacetime in terms of the dS plane waves, and allow us to bring the general solution (\ref{=/2/5}) into the intrinsic form,
\begin{eqnarray}\label{intrinsic}
h_{\mu\nu} \equiv {\cal E}_{\mu\nu} \Phi^{-3}_{Llm}(\rho,u) = (x^\alpha_\mu x^\beta_\nu {\cal E}_{\alpha\beta})\Phi^{-3}_{Llm}(\rho,u).
\end{eqnarray}

According to our choice of the global coordinate system, the dS-invariant inner product on the space of solutions (\ref{intrinsic}) is defined by \cite{Gazeau329}
\begin{eqnarray} \label{B1}
&\langle h_1,h_2\rangle = \frac{i}{H^2} \int_{S^3,\rho=0} [({h_1})^\ast\cdot\cdot\partial_\rho{h_2} &\nonumber\\
&- 2c ((\partial_\rho x)\cdot{({h_1})}^\ast)\cdot(\partial\cdot{h_2}) - (1^\ast \leftrightharpoons 2)]d\sigma(u),&
\end{eqnarray}
where $h_1$ and $h_2$ are two arbitrary modes. Let us take a closer look at the behavior of this inner product. With respect to the following identity for hypergeometric functions \cite{Magnus}
\begin{eqnarray}
_2F_1(a,b;c;z) = {(1-z)^{(c-a-b)}} {_2F_1}(c-a,c-b;c;z),\nonumber
\end{eqnarray}
one can find the alternative form of (\ref{SOF}) as
\begin{eqnarray}
\Phi_{Llm}^{-3}(x) = i^{L+3} e^{-iL\rho}  \frac{\Gamma(L+3)}{(L+1)! \Gamma(3)}\hspace{2cm} \nonumber\\
\times {_2F_1}(-1,L;L+2;-e^{-2i\rho})Y_{Llm}(u).
\end{eqnarray}
Now, considering the behavior of the hypergeometric functions \cite{Magnus},
$${_2F_1}(-1,L;L+2;-e^{-2i\rho})= 1-\frac{L}{L+2}e^{-2i\rho},$$
it is clear that we face a degeneracy in the set of modes for $L\geq0$; the mode associated with $L=0$ (i.e., ${\cal E}_{\mu\nu} \Phi^{-3}_{0,0,0}$ in which $\Phi^{-3}_{0,0,0}$ is a constant function) is orthogonal to the entire set of modes including itself. The appearance of this degeneracy raises a natural question, which is decisive in evaluating different approaches to the long-standing puzzle of dS breaking in linearized quantum gravity: does this degeneracy reflect a hitherto `hidden' symmetry of the existing physics?

Thanks to the mathematical structure presented thus far based on the ambient space formalism, one can easily observe that the theory, besides the spacetime symmetries generated by the Killing vectors and the gauge transformation (\ref{GT}), is also invariant under a `hidden' gauge-like symmetry, i.e., for any constant function $\chi$ and quite independent of whatever gauge fixing parameter $c$ that one chooses, the inner product and equation of motion are invariant under the symmetry
\begin{eqnarray}\label{g-like}
h_{\mu\nu}\rightarrow h_{\mu\nu} + {\cal E}_{\mu\nu}\chi.
\end{eqnarray}
This proves that the invariant null-norm subspace, generated by $L=0$, can be interpreted as a space of `gauge' states in the set of solutions. As a direct consequence, this argument explicitly casts serious doubts on the viability of the appealing picture of dS quantum gravity in the Bunch-Davies vacuum when it is evaluated in the TTS gauge ($L\geq2$), and shows that this quantization scheme, ignoring the $L=0$ mode in the set of solutions and consequently the symmetry (\ref{g-like}) reflected by it, does not transform correctly under the whole symmetries of the theory. Therefore, even obtaining an infrared finite graviton propagator in the TTS gauge (see \cite{higuchi,Bernar,HH Higuchi'',HH Frob,HH Bernar,Fewster,Faizal&Higuchi,HH Higuchi',Marolf&Morrison,HH Higuchi,higuchi1,HH Hartle,HH Allen,HH Marolf,FrobJCAP}) has no physical meaning, because it is not covariant anyway.

Now, to get our main result, we need to evaluate the behavior of the local symmetry (\ref{g-like}) of the classical theory at the quantum level. The presence of this classical symmetry compels us to delve more deeply into the $L=0$ mode in order to circumvent the degeneracy problem associated with it. [Recall that, the starting point toward a canonical quantization would be to build a complete, non-degenerate and fully invariant space of solutions to the field equation.] In this regard, by solving the field equation directly for $L=0$, more accurately Eq. (\ref{mmcsf}) (see \cite{Gazeau1415}), we obtain two-independent solutions (including the gauge-like solutions appeared in (\ref{g-like})) as follows
\begin{eqnarray}
\underline{h}_{\mu\nu}^{(0,0,0)} = {\cal E}_{\mu\nu}\Big(\frac{H}{2\pi}\Big),\; \underline{\underline{h}}_{\mu\nu}^{(0,0,0)} = {\cal E}_{\mu\nu}\Big( -i\frac{H}{2\pi}[\rho+\frac{1}{2}\sin 2\rho]\Big).\nonumber
\end{eqnarray}
Both modes however have null norm, and the constant factors are selected to get $\langle \underline{h}_{\mu\nu}^{(0,0,0)} , \underline{\underline{h}}_{\mu\nu}^{(0,0,0)} \rangle =1$. Therefore, to circumvent the degeneracy problem, we define the `true' normalized zero mode of the system as following combination of $\underline{h}_{\mu\nu}^{(0,0,0)}$ and ${\underline{\underline{h}}_{\mu\nu}^{(0,0,0)}}$,
\begin{eqnarray}\label{newmode}
h_{\mu\nu}^{(0,0,0)} = {\cal E}_{\mu\nu} \Phi^{-3}_{0,0,0}= {\cal E}_{\mu\nu} \Big(\frac{H}{2\pi} - i\frac{H}{4\pi}[\rho+\frac{1}{2}\sin 2\rho]\Big),\;
\end{eqnarray}
with $\langle h_{\mu\nu}^{(0,0,0)},h_{\mu\nu}^{(0,0,0)}\rangle =1$. Taking into account this new mode (\ref{newmode}) interestingly gives a complete set of strictly positive-norm modes $h_{\mu\nu}^{(Llm)}$ for $L\geq 0$. The space spanned by these modes, however, is not dS invariant, whose root is indeed the modified zero mode (again, the gauge-like symmetry (\ref{g-like}) unavoidably compels us to consider this mode). More accurately, applying the dS boost generators on this mode produces all the negative and positive norm modes of the field equation. For instance, considering the action of the dS boost generator $(L_{03} + iL_{04})$, we have
\begin{eqnarray}\label{action}
(L_{03} + iL_{04})h_{\mu\nu}^{(0,0,0)}= \Big((L_{03} + iL_{04}){\cal E}_{\mu\nu}\Big)\Phi^{-3}_{0,0,0} \hspace{0.4cm}\nonumber\\
+ {\cal E}_{\mu\nu} \Big((M_{03} + iM_{04})\Phi^{-3}_{0,0,0}\Big).
\end{eqnarray}
The first term is trivially invariant under the dS group action. The dS invariance, however, is broken because of the second term. In fact, a direct computation gives
\begin{eqnarray}
\mbox{(\ref{action})}= \mbox{`invariant terms'}\hspace{5cm} \nonumber\\
- \frac{\sqrt{6}}{4} \Big(h_{\mu\nu}^{(1,1,0)} + ih_{\mu\nu}^{(1,0,0)} + (1+i)(h_{\mu\nu}^{(1,0,0)})^\ast \Big),\nonumber
\end{eqnarray}
where $(h_{\mu\nu}^{(1,0,0)})^\ast$ stands for complex conjugates of $h_{\mu\nu}^{(1,0,0)}$, with $\langle (h_{\mu\nu}^{(1,0,0)})^\ast, (h_{\mu\nu}^{(1,0,0)})^\ast \rangle= -1$. [See Ref. \cite{Gazeau1415} for the action of $(M_{03} + iM_{04})$ on the scalar structure function $\Phi^{-3}_{0,0,0}$.]

This result explicitly reveals that the dS graviton quantum field constructed through canonical quantization and the usual representation of the canonical commutation relations suffers from an anomalous gauge-like symmetry (the local symmetry of the classical theory, see (\ref{g-like}), is absent at the quantum level), for which, one has to deal with the propagation of un-physical negative norm states in the theory. As we have shown above, the appearance of this anomaly is $c$ independent, and therefore, cannot be gauged away.

\section{Discussion}
We here recall that, gauge anomalies are related to the global topology and have the common feature \cite{42} that the \emph{Gauss constraint} cannot anymore be implemented on the physical states \cite{13}. The reason is that the anomalous Ward identity relates the time evolution of the Gauss constraint to the conservation law for the matter current on the spacelike hypersurface \cite{23}. In this sense, the uncovered local gauge anomaly associated with the classical symmetry (\ref{g-like}) would be catastrophic for a consistent quantum field theory of gravity in dS spacetime and must be cancelled at all costs.

The cancelation of this local anomaly clearly poses strong restrictions on the theory. In fact, with respect to canonical quantization and the usual representation of the canonical commutation relations, the only possible ways out seem to weakening the covariance condition by dropping the dS boost invariance in a way or another: considering vacua invariant under a subgroup of the dS isometry group only, the so-called spontaneous symmetry breaking, for example one can easily prove that the above set of solutions for $L\geq0$ is $SO(4)$ invariant; restricting the field to a subset of the dS manifold; considering invariance with respect to the Lie algebra of the dS group rather than under the full group action.

Frankly speaking, the above argument explicitly reveals that the long-standing puzzle of dS symmetry breaking in linearized quantum gravity is indeed a matter of gauge anomaly. Precisely, with respect to the usual representation of the canonical commutation relations, this symmetry breaking is the price that must be paid in order to cancel the local anomaly associated with (\ref{g-like}) and to preserve the consistency of the theory.

Here, we must underline that, as dS symmetries are basic symmetries of field dynamics in dS space, one also might treat this anomaly, that is, a major stumbling block in preserving the conservation laws, through the idea of changing the quantization scheme in such a way that respects the dS symmetries. Evidence in favour of this idea was presented in \cite{Balasubramanian} by studying the dS/CFT correspondence. In this paper, regarding the fundamental difficulties of defining a measure of mass, angular momentum and other conserved charges in dS spacetime, the authors have calculated the Brown-York boundary stress tensor of asymptotically dS spacetimes and have considered it to characterized a novel notion of mass and conserved charges. They have correctly claimed that the obtained quantities would be the stress tensor and charges of the dual theory, if such a theory were defined for dS in a manner analogous to the AdS/CFT correspondence. The authors have remarkably argued that the natural conjecture is that admitting a non-canonical Hilbert space structure would result in such a dual to Lorentzian dS space.

Further evidence appeared in \cite{CCP} by studying a potential relevance between the Krein-Gupta-Bleuler (KGB) vacuum leading to a fully covariant quantum field theory for gravity in dS space and the observable smallness of the cosmological constant. The KGB quantization scheme, possessing a rigorous group theoretical content which provides a well-defined meaning of massive and massless dS fields (see \cite{Gazeau1415} and references therein), is a new canonical quantization method based on weaker conditions, which do not prohibit negative norm states in the definition of the field. In this sense, it provides a unified framework to treat gauge and gauge-like quantum field theories in dS space (see \cite{Bamba2,in preparation} and references therein). In the KGB context, despite the presence of negative norm sates in the theory, the energy operator is positive in all physical states, and vanishes in the vacuum. Recently, a preliminary estimate of the expected order of magnitude of vacuum energy density stored in the cosmological constant today with respect to the KGB vacuum, presented in \cite{CCP}, has been shown that the obtained result demonstrates a remarkable coincidence with the observational data. This coincidence seems to be almost too good to be just an accident.

Finally, we would like to emphasize that linearized gravity seems to be quite special and non-universal in its features, then it is very tempting to adopt the idea of changing the quantization scheme as a working hypothesis. Of course, there is certainly a tremendous amount of work still to be done in order to single out which of these physically relevant approaches (including the usual canonical quantization ones with the restrictive versions of the dS covariance and the non-standard quantization schemes with the full dS covariance) can be actually the viable one.

\section*{Acknowledgements}
This work was partially supported by the JSPS KAKENHI Grant Number JP 25800136 and Competitive Research Funds for Fukushima University Faculty (18RI009), and the National Natural Science Foundation of China with the Grants Nos: 11375153 and 11675145.


\begin{thebibliography}{a}

\bibitem{Antoniadis2007} I. Antoniadis, P.O. Mazur, and E. Mottola, New Jour. of phys. 9, 11 (2007).

\bibitem{AF} B. Allen and A. Folacci, Phys. Rev. D 35, 3771 (1987).

\bibitem{MorrisonarXiv} I.A. Morrison, \emph{On cosmic hair and ``de Sitter breaking" in linearized quantum gravity}, arXiv:1302.1860 [gr-qc].

\bibitem{higuchi} A. Higuchi, Class. Quant. Grav. 8, 2005 (1991).

\bibitem{Bernar} R.P. Bernar, L.C.B. Crispino, and A. Higuchi, Phys. Rev. D 97, 085005 (2018).

\bibitem{HH Higuchi''} A. Higuchi, and N. Rendell, Class. Quant. Grav. 35, 115004 (2018).

\bibitem{HH Frob} M.B. Frob, A. Higuchi, and W.C.C. Lima, Phys. Rev. D 93, 124006 (2016).

\bibitem{HH Bernar} R.P. Bernar, L.C.B. Crispino, and A. Higuchi, Phys. Rev. D 90, 024045 (2014).

\bibitem{FrobJCAP} M.B. Frob, JCAP 1412, no.12, 010 (2014).

\bibitem{Fewster} C.J. Fewster, D.S. Hunt, Rev. Math. Phys. 25, 1330003 (2013).

\bibitem{Faizal&Higuchi} M. Faizal, and A. Higuchi, Phys. Rev. D 85, 124021 (2012).

\bibitem{HH Higuchi'} A. Higuchi, D. Marolf, and I.A. Morrison, Class. Quant. Grav. 28, 245012 (2011).

\bibitem{Marolf&Morrison} D. Marolf, and I.A. Morrison, Class. Quantum Grav. 26, 235003 (2009).

\bibitem{HH Higuchi} A. Higuchi, and R.H. Weeks, Class. Quant. Grav. 20, 3005 (2003).

\bibitem{higuchi1} A. Higuchi, and S.S. Kouris, Class. Quant. Grav. 18, 4317 (2001).

\bibitem{HH Hartle} J.B. Hartle, and D. Marolf, Phys. Rev. D 56, 6247 (1997).

\bibitem{HH Allen} B. Allen, Phys. Rev. D 51, 5491 (1995).

\bibitem{HH Marolf} D. Marolf, Class. Quant. Grav. 12, 2469 (1995).

\bibitem{Dixmier} J. Dixmier, Bull. Soc. Math. Fr. 89, 9 (1961).

\bibitem{Gazeau2533} J.P. Gazeau, M. Hans, J. Math. Phys. 29, 2533 (1988).

\bibitem{levy} M. Levy-Nahas, J. Math. Phys. 8, 1211 (1967).

\bibitem{barut} A.O. Barut , A. B\"ohm, J. Math. Phys. 11, 2938 (1970).

\bibitem{bacry} H. Bacry, J.M. Levy-Leblond, J. Math. Phys. 9, 1605 (1968).

\bibitem{Fronsdal1979} C. Fronsdal, Phys. Rev. D 20, 848 (1979).

\bibitem{PejhanII} H. Pejhan and S. Rahbardehghan, Phys. Rev. D 94, 104030 (2016).

\bibitem{Bamba2} H. Pejhan, K. Bamba, S. Rahbardehghan, and M. Enayati, Phys. Rev. D 98, 045007 (2018).

\bibitem{Gazeau1984} J.P. Gazeau, Lett. Math. Phys. 8, 507 (1984).

\bibitem{Gazeau1985g} J.P. Gazeau, J. Math. Phys. 26, 1847 (1985).

\bibitem{BGM} J. Bros, J. P. Gazeau, and U. Moschella, Phys. Rev. Lett. 73, 1746 (1994).

\bibitem{BM} J. Bros and U. Moschella, Rev. Math. Phys. 8, 327 (1996).

\bibitem{MVFgazeau} T. Garidi, J.P. Gazeau, S. Rouhani, and M. V. Takook, J. Math. Phys. 49, 032501 (2008).

\bibitem{Gazeau329} J.P. Gazeau, M. Hans, and R. Murenzi, Class. Quant. Grav. 6, 329 (1989).

\bibitem{Magnus} W. Magnus, F. Oberhettinger, R.P. Soni, \emph{Formulas and theorems for the special functions of mathematical physics}, Springer-Verlag, New York, 1966.

\bibitem{Gazeau1415} J.P. Gazeau, J. Renaud, and M.V. Takook, Class. Quant. Grav. 17, 1415 (2000).

\bibitem{42} P.h. Nelson, and L. Alvarez–Gaum$\mbox{\'{e}}$, Commun. Math. Phys. 99, 103 (1985).

\bibitem{13} L.D. Faddeev, Phys. Lett. B 145, 81 (1984).

\bibitem{23} W. Jiang, J. Math. Phys. 32, 3409 (1991).

\bibitem{Balasubramanian} V. Balasubramanian, J. de Boer, D. Minic, Phys. Rev. D 65, 123508 (2002).

\bibitem{CCP} H. Pejhan, K. Bamba, M. Enayati, and S. Rahbardehghan, Phys. Lett. B 785, 567 (2018).

\bibitem{in preparation} H. Pejhan, M. Enayati, J.P. Gazeau, and A. Wang, \emph{Gupta-Bleuler quantization for linearized gravity in de Sitter spacetime}, arXiv:1906.06644 (2019).

\end{thebibliography}
\end{document}